\documentclass[fleqn]{mat01}
\usepackage{times,mathtimy,amssymb,epsfig}
\begin{document}

\setcounter{page}{451}
\firstpage{451}

\newtheorem{theore}{Theorem}
\renewcommand\thetheore{\arabic{section}.\arabic{theore}}
\newtheorem{theor}[theore]{\bf Theorem}

\def\thoe{\trivlist\item[\hskip\labelsep{{\bf Theorem}}]}
\newtheorem{theo}{\bf Theorem}
\renewcommand\thetheo{\arabic{theo}}
\newtheorem{propo}[theo]{\rm PROPOSITION}
\newtheorem{coro}[theo]{\rm COROLLARY}
\newtheorem{lem}[theo]{Lemma}
\newtheorem{fact}[theo]{Fact}
\newtheorem{claim}{Claim}
\newtheorem{rem}[theo]{Remark}
\newtheorem{definit}[theo]{\rm DEFINITION}
\newtheorem{exampl}{Example}

\renewcommand{\theequation}{\thesection\arabic{equation}}

\title{On Howard's conjecture in heterogeneous shear flow problem}

\markboth{R~G~Shandil and Jagjit Singh}{Linear heterogeneous shear flow
instability}

\author{R~G~SHANDIL and JAGJIT SINGH$^{*}$}

\address{Department of Mathematics, H.P. University, Shimla~171~005, India\\
\noindent $^{*}$Sidharth Govt. Degree College, Nadaun, Dist. Hamirpur~177~033, India}

\volume{113}

\mon{November}

\parts{4}


\begin{abstract}
Howard's conjecture, which states that in the linear instability problem
of inviscid heterogeneous parallel shear flow growth rate of an
arbitrary unstable wave must approach zero as the wave length decreases
to zero, is established in a mathematically rigorous fashion for plane
parallel heterogeneous shear flows with negligible buoyancy force
$g\beta \ll 1$ (Miles~J~W, {\it J. Fluid Mech.} {\bf 10} (1961)
496--508), where $\beta$ is the basic heterogeneity distribution
function).
\end{abstract}

\maketitle

\section{Introduction}

In the stability problem of inviscid heterogeneous parallel shear flows,
Howard \cite{4}, making use of a novel transformation of the dependent
variable was not only able to prove the validity of Taylor's conjecture
without the restrictive conditions under which Miles \cite{5} got the
result but he was also able to obtain a semicircular region in the
$c_{r} c_{i}$-plane in which the complex velocity of an unstable wave
must lie. Howard also made a conjectural assertion to the effect that
the growth rate of an arbitrary unstable mode must approach zero as the
wave number approaches infinity. This conjecture of Howard has also
drawn the attention of researchers. Banerjee {\it et~al} \cite{1} were
able to validate the correctness of this conjecture for the special case
of inviscid homogeneous parallel shear flows. Their approach consisted
of combining the governing equations and boundary conditions in an
innovative way and thereby deriving an upper bound of the growth rate
under consideration. Banerjee {\it et~al} \cite{2} attempted Howard's
conjecture in heterogeneous parallel shear flows but succeeded only in
proving it in the case of the Garcia-type \cite{3} flows wherein the
basic velocity distribution has a point of inflexion in the domain of
the flow while the vertical velocity gradient of the basic density
distribution vanishes at this point.

In this paper we prove, in a mathematically rigorous fashion, that the
growth rate of an arbitrary unstable wave must approach zero as the wave
number approaches infinity in the linear instability of non-viscous
heterogeneous parallel shear flows with negligible buoyancy force
($g\beta \ll 1$, \cite{5}) so that second- and higher-order terms in
$g\beta$, where $\beta$ is the basic heterogeneity distribution
function, are negligible as compared to the first-order terms in
$g\beta$.

\section{Mathematical formulation of the problem}

The basic equations governing the linear instability in a Boussinesq
inviscid parallel shear flow which is confined between two rigid
horizontal boundaries is given by

\pagebreak

$\left.\right.$\vspace{-2.3pc}

\begin{equation}
(D^{2} - \alpha^{2}) w - \left( \frac{U''}{U - c} \right) w + \left(
\frac{g\beta}{(U - c)^{2}} \right) w = 0,
\end{equation}
where $z \in [z_{1}, z_{2}]$ is the real independent variable and stands
for the vertical coordinate, $D \equiv {\rm d}/{\rm d}z, U(z)$ is a
twice continuously differentiable function of $z$ and stands for the
basic velocity distribution, $\beta(z)$ is a non-negative continuous
function of $z$ and stands for the basic heterogeneity distribution,
$w(z)$ is the dependent variable and stands for the $z$-component of the
perturbation velocity, $c = c_{r} + ic_{i}$ is a complex constant in
general and stands for the complex wave velocity of the perturbation
wave with $c_{r}$ as the phase velocity and $c_{i}$ as the amplification
factor, and $\alpha^{2}$ is a positive constant which satisfies $0 <
\alpha^{2} < \infty$ and stands for the square of the wave number of the
perturbation wave. The boundary conditions associated with the problem
are that $w(z)$ vanishes on the rigid horizontal boundaries at $z =
z_{1}$ and $z = z_{2}$ i.e.,
\begin{equation}
w(z_{1}) = w(z_{2}) = 0.
\end{equation}
(The boundaries in the limiting case may recede to $\pm \infty$.)

For the existence of a non-trivial solution of eqs~(2.1) and (2.2) we
have a double eigenvalue problem for the determination of $c_{r}$ and
$c_{i}$ for the prescribed values of $\alpha^{2}$ and the flow is
unstable if such solutions exist for which the imaginary part $c_{i}$ of
$c$ is greater than zero.

\section{Mathematical analysis}

Firstly, we prove the following two lemmas:

\begin{lem} A necessary condition for the existence of a non-trivial
solution $(w, c, \alpha^{2})$ with $c_{i} > 0$ of eqs~{\rm (2.1)} and
{\rm (2.2)} is that the integral relations
\setcounter{equation}{0}
\begin{align}
\int_{z_{1}}^{z_{2}} (\vert Dw\vert^{2} + \alpha^{2} |w|^{2}) {\rm d}z &+
\int_{z_{1}}^{z_{2}} \frac{U'' (U - c_{r})}{(U - c_{r})^{2} + c_{i}^{2}}
|w|^{2} {\rm d}z\nonumber\\[.2pc]
&- \int_{z_{1}}^{z_{2}} \frac{g\beta \{ (U - c_{r})^{2} -
c_{i}^{2}\}}{\{(U - c_{r})^{2} + c_{i}^{2}\}^{2}} |w|^{2} = 0
\end{align}
and
\begin{equation}
\int_{z_{1}}^{z_{2}} \frac{U''}{(U - c_{r})^{2} + c_{i}^{2}} |w|^{2} {\rm
d}z - 2 \int_{z_{1}}^{z_{2}} \frac{g\beta(U - c_{r})}{\{(U - c_{r})^{2}
+ c_{i}^{2}\}^{2}} |w|^{2} = 0,
\end{equation}
are true.
\end{lem}

\begin{proof}
We multiply eq.~(2.1) by $w^{*}$ (the complex conjugate of $w$)
throughout and integrate the resulting equation over the domain of $z$,
to get
\begin{align}
&\int_{z_{1}}^{z_{2}}w^{*} (D^{2} - \alpha^{2}) w{\rm d}z -
\int_{z_{1}}^{z_{2}} w^{*} \left( \frac{U''}{U - c} \right) w {\rm d}z\nonumber\\[.2pc]
&\quad\ + \int_{z_{1}}^{z_{2}} w^{*} \left( \frac{g\beta}{(U -
c)^{2}}\right) w {\rm d}z = 0.
\end{align}

In order to calculate the first term of the first integral on the left
hand side of eq.~(3.3), we integrate it by parts once and use the
boundary condition (2.2), to derive
\begin{equation}
\int_{z_{1}}^{z_{2}} (|Dw|^{2} + \alpha^{2} |w|^{2}) {\rm d}z +
\int_{z_{1}}^{z_{2}} \left( \frac{U''}{U - c} - \frac{g\beta}{(U -
c)^{2}}\right) |w|^{2} {\rm d}z = 0.
\end{equation}
Now equating real and imaginary parts of the two sides of eq.~(3.4) and
cancelling $c_{i}(> 0)$ throughout from the imaginary part, we get
\begin{align}
\int_{z_{1}}^{z_{2}} (|Dw|^{2} + \alpha^{2} |w|^{2}) {\rm d}z &+
\int_{z_{1}}^{z_{2}} \frac{U'' (U - c_{r})}{(U - c_{r})^{2} +
c_{i}^{2}}|w|^{2}{\rm d}z\nonumber\\[.2pc]
&- \int_{z_{1}}^{z_{2}}\frac{g\beta\{ (U -
c_{r})^{2} - c_{i}^{2}\}}{\{(U - c_{r})^{2} + c_{i}^{2}\}^{2}} |w|^{2} =
0
\end{align}
and
\begin{equation}
\int_{z_{1}}^{z_{2}} \frac{U''}{(U - c_{r})^{2} + c_{i}^{2}} |w|^{2}
{\rm d}z - 2 \int_{z_{1}}^{z_{2}}\frac{g\beta (U - c_{r})}{\{(U -
c_{r})^{2} + c_{i}^{2}\}^{2}} |w|^{2} = 0,
\end{equation}
and this proves the lemma.
\end{proof}

\begin{lem} A necessary condition for the existence of a non-trivial
solution $(w, c, \alpha^{2})$ with $c_{i} > 0$ of eqs~{\rm (2.1)} and
{\rm (2.2)} is that the integral relation
\begin{align}
&\int_{z_{1}}^{z_{2}} (|D^{2}w|^{2} + \alpha^{2} |Dw|^{2}) {\rm d}z -
\alpha^{2} \int_{z_{1}}^{z_{2}} \frac{U'' (U - c_{r})}{(U - c_{r})^{2} +
c_{i}^{2}}|w|^{2}{\rm d}z\nonumber\\[.3pc]
&\quad\ + \alpha^{2} \int_{z_{1}}^{z_{2}}\frac{g\beta\{ (U -
c_{r})^{2} - c_{i}^{2}\}}{\{(U - c_{r})^{2} + c_{i}^{2}\}^{2}} |w|^{2}
{\rm d}z  - \int_{z_{1}}^{z_{2}} \frac{(U'')^{2}}{(U - c_{r})^{2} +
c_{i}^{2}}|w|^{2}{\rm d}z\nonumber\\[.3pc]
&\quad\ + 2 \int_{z_{1}}^{z_{2}} \frac{g\beta(U -
c_{r})}{\{(U - c_{r})^{2} + c_{i}^{2}\}^{2}} |w|^{2} {\rm d}z - \int_{z_{1}}^{z_{2}}
\frac{g^{2}\beta^{2}}{\{(U - c_{r})^{2} + c_{i}^{2}\}^{2}} |w|^{2} {\rm
d}z = 0
\end{align}
and
\begin{align}
&\int_{z_{1}}^{z_{2}} (|D^{2}w|^{2} + 2\alpha^{2} |Dw|^{2} + \alpha^{4}
|w|^{2}) {\rm d}z - \int_{z_{1}}^{z_{2}} \frac{(U'')^{2}}{\{(U -
c_{r})^{2} + c_{i}^{2}\}}|w|^{2}{\rm d}z\nonumber\\[.3pc]
&\quad\ + 2\int_{z_{1}}^{z_{2}} \frac{g\beta U'' ((U -
c_{r}))}{\{(U - c_{r})^{2} + c_{i}^{2}\}^{2}} |w|^{2} {\rm d}z
-\! \int_{z_{1}}^{z_{2}} \frac{g^{2}\beta^{2}}{\{(U - c_{r})^{2} +
c_{i}^{2}\}^{2}} |w|^{2} {\rm d}z = 0
\end{align}
are true.
\end{lem}

\begin{proof}
We multiply eq.~(2.1) by $D^{2} w^{*}$ (the complex conjugate of $D^{2}
w$) throughout and integrate the resulting equation over the domain of
$z$, to get
\begin{align}
\int_{z_{1}}^{z_{2}} D^{2} w^{*} (D^{2} - \alpha^{2})w {\rm d}z &-
\int_{z_{1}}^{z_{2}} D^{2} w^{*} \left( \frac{U''}{U - c} \right) w{\rm
d}z\nonumber\\[.3pc]
&+ \int_{z_{1}}^{z_{2}} D^{2} w^{*} \left( \frac{g\beta}{(U -
c)^{2}} \right) w{\rm d}z = 0.
\end{align}

In order to calculate the first term of the first integral on the left
hand side of eq.~(3.9), we integrate it by parts once and use the
boundary condition (2.2), to derive
\begin{equation}
\int_{z_{1}}^{z_{2}} (|D^{2} w|^{2} + \alpha^{2} |Dw|^{2}) {\rm d}z -
\int_{z_{1}}^{z_{2}} D^{2} w^{*} \left( \frac{U''}{U - c} -
\frac{g\beta}{(U - c)^{2}} \right) w{\rm d}z = 0.
\end{equation}
Now from eq.~(2.1), we have
\begin{equation}
D^{2} w = \alpha^{2} w + \left( \frac{U''}{U -c} \right) w - \left(
\frac{g\beta}{(U - c)^{2}} \right) w.
\end{equation}
Taking the complex conjugate of both sides of eq.~(3.11), we derive
\begin{equation}
D^{2} w^{*} = \alpha^{2} w^{*} + \left( \frac{U''}{U - c^{*}} \right)
w^{*} - \left( \frac{g\beta}{(U - c^{*})^{2}} \right) w^{*}.
\end{equation}
Substituting the value of $D^{2} w^{*}$ from eq.~(3.12) into eq.~(3.10),
we derive
\begin{align}
\int_{z_{1}}^{z_{2}} (|D^{2} w|^{2} &+ \alpha^{2} |Dw|^{2}) {\rm d}z -
\int_{z_{1}}^{z_{2}} \left( \alpha^{2} + \frac{U''}{(U - c^{*})} -
\frac{g\beta}{(U - c^{*})^{2}} \right) \nonumber\\[.2pc]
&\times w^{*} \left( \frac{U''}{U - c} - \frac{g\beta}{(U -
c)^{2}} \right) w{\rm d}z = 0.
\end{align}
On simplification, eq.~(3.13) can be written in the form
\begin{align}
\int_{z_{1}}^{z_{2}} (|D^{2} w|^{2} &+ \alpha^{2} |Dw|^{2}) {\rm d}z -
\alpha^{2} \int_{z_{1}}^{z_{2}} \frac{U'' (U - c_{r} + ic_{i})}{(U -
c_{r})^{2} + c_{i}^{2}} |w|^{2} {\rm d}z\nonumber\\[.3pc]
&- \int_{z_{1}}^{z_{2}}
\frac{(U'')^{2}}{\{(U - c_{r})^{2} + c_{i}^{2}\}} |w|^{2} {\rm
d}z\nonumber\\[.3pc]
&+ \alpha^{2} \int_{z_{1}}^{z^{2}} \frac{g\beta \{(U - c_{r})^{2} -
c_{i}^{2} + 2ic_{i} (U - c_{r})\}}{\{(U - c_{r})^{2} + c_{i}^{2} \}^{2}}
|w|^{2} {\rm d}z\nonumber\\[.3pc]
&+ \int_{z_{1}}^{z_{2}} \frac{g\beta U'' ((U - c_{r}) -
ic_{i})}{\{ (U - c_{r})^{2} + c_{i}^{2}\}^{2}} |w|^{2} {\rm d}z\nonumber\\[.3pc]
&+ \int_{z_{1}}^{z_{2}} \frac{g\beta U'' \{(U - c_{r}) +
ic_{i}\}}{\{ (U - c_{r})^{2} + c_{i}^{2}\}^{2}} |w|^{2} {\rm d}z\nonumber\\[.3pc]
&+ \int_{z_{1}}^{z_{2}} \frac{g^{2}\beta^{2}}{\{ (U - c_{r})^{2}
+ c_{i}^{2}\}^{2}} |w|^{2} {\rm d}z = 0.
\end{align}

Equating the real parts on both sides of eq.~(3.14) we obtain
\begin{align}
&\int_{z_{1}}^{z_{2}} (|D^{2} w|^{2} + \alpha^{2} |Dw|^{2}) {\rm d}z -
\alpha^{2} \int_{z_{1}}^{z_{2}} \frac{U'' (U - c_{r})}{(U -
c_{r})^{2} + c_{i}^{2}} |w|^{2} {\rm d}z\nonumber\\[.3pc]
&\quad\ - \int_{z_{1}}^{z_{2}}
\frac{(U'')^{2}}{\{(U - c_{r})^{2} + c_{i}^{2}\}} |w|^{2} {\rm
d}z + \alpha^{2} \int_{z_{1}}^{z_{2}} \frac{g\beta \{(U - c_{r})^{2} -
c_{i}^{2}\}}{\{(U - c_{r})^{2} + c_{i}^{2} \}^{2}}
|w|^{2} {\rm d}z\nonumber\\[.3pc]
&\quad\ + 2\int_{z_{1}}^{z_{2}} \frac{g\beta U'' ((U - c_{r}))}{\{ (U -
c_{r})^{2} + c_{i}^{2}\}^{2}} |w|^{2} {\rm d}z \!-\!
\int_{z_{1}}^{z_{2}} \frac{g^{2}\beta^{2}}{\{ (U - c_{r})^{2}
+ c_{i}^{2}\}^{2}} |w|^{2} {\rm d}z = 0.
\end{align}
Multiplying eq.~(3.1) by $\alpha^{2}$ and adding to eq.~(3.15) we get
\begin{align}
&\int_{z_{1}}^{z_{2}} (|D^{2} w|^{2} + 2\alpha^{2} |Dw|^{2} + \alpha^{4}
|w|^{2}) {\rm d}z - \int_{z_{1}}^{z_{2}} \frac{(U'')^{2}}{\{(U -
c_{r})^{2} + c_{i}^{2}\}} |w|^{2} {\rm d}z\nonumber\\[.3pc]
&\quad\ + 2\int_{z_{1}}^{z_{2}} \frac{g\beta U'' ((U - c_{r}))}{\{ (U -
c_{r})^{2} + c_{i}^{2}\}^{2}} |w|^{2} {\rm d}z \!-\! \int_{z_{1}}^{z_{2}}
\frac{g^{2}\beta^{2}}{\{ (U - c_{r})^{2} + c_{i}^{2}\}^{2}} |w|^{2} {\rm
d}z = 0,
\end{align}
which completes the proof of the lemma.
\end{proof}

\setcounter{section}{5}
\setcounter{theo}{0}
\begin{theor}[\!] If $(w, c, \alpha^{2})$ is a non-trivial solution of
eqs~{\rm (2.1)} and {\rm (2.2)} with $c_{i} > 0$ and heterogeneity
factor is small so that second- and higher-order terms in $g \beta${\rm ,} are
negligible as compared to the first-order terms in $g\beta$ then
$\alpha c_{i} \rightarrow 0$ as $\alpha \rightarrow \infty$.
\end{theor}

\setcounter{section}{3}

\begin{proof}
Since heterogeneity factor is small so that second- and higher-order
terms in $g\beta$, where $\beta$ is the basic heterogeneity distribution
function, are negligible as compared to the first-order terms in
$g\beta$, therefore eq.~(3.8) reduces to
\begin{align}
\int_{z_{1}}^{z_{2}} (|D^{2} w|^{2} &+ 2\alpha^{2} |Dw|^{2} + \alpha^{4}
|w|^{2}) {\rm d}z - \int_{z_{1}}^{z_{2}} \frac{(U'')^{2}}{\{(U -
c_{r})^{2} + c_{i}^{2}\}} |w|^{2} {\rm d}z\nonumber\\[.3pc]
&+ 2\int_{z_{1}}^{z_{2}} \frac{g\beta U'' ((U - c_{r}))}{\{ (U -
c_{r})^{2} + c_{i}^{2}\}^{2}} |w|^{2} {\rm d}z = 0.
\end{align}
Since $c_{i} > 0$ and $1/\{(U - c_{r})^{2} + c_{i}^{2}\} \leq
1/c_{i}^{2}$, therefore, from eq.~(3.17) we derive that
\begin{align}
\int_{z_{1}}^{z_{2}} (|D^{2} w|^{2} &+ 2\alpha^{2} |Dw|^{2} + \alpha^{4}
|w|^{2}) {\rm d}z - \int_{z_{1}}^{z_{2}} \frac{[(U'')^{2}
]_{\max}}{c_{i}^{2}} |w|^{2} {\rm d}z\nonumber\\[.3pc]
&+ 2 \int_{z_{1}}^{z_{2}} \frac{g\beta U'' (U - c_{r})c_{i}}{\{
(U - c_{r})^{2} + c_{i}^{2}\}^{2} c_{i}} |w|^{2} {\rm d}z \leq 0.
\end{align}
where $[(U'')^{2}]_{\max}$ stands for the
maximum value of the bracketed expression over the interval $[z_{1},
z_{2}]$. Further, since $(U - c_{r})^{2} + c_{i}^{2} \geq 2 (U - c_{r})
c_{i}$, therefore from inequality (3.18) we derive that

\pagebreak

$\left.\right.$\vspace{-2.5pc}

\begin{align}
\int_{z_{1}}^{z_{2}} (|D^{2} w|^{2} &+ 2\alpha^{2} |Dw|^{2} + \alpha^{4}
|w|^{2}) {\rm d}z - \int_{z_{1}}^{z_{2}} \frac{[(U'')^{2}
]_{\max}}{c_{i}^{2}} |w|^{2} {\rm d}z\nonumber\\[.2pc]
&- \int_{z_{1}}^{z_{2}} \frac{g\beta |U''| ((U - c_{r})^{2} +
c_{i}^{2})}{\{ (U - c_{r})^{2} + c_{i}^{2}\}^{2} c_{i}} |w|^{2} {\rm d}z
\leq 0,
\end{align}
or
\begin{align}
\int_{z_{1}}^{z_{2}} (|D^{2} w|^{2} &+ 2\alpha^{2} |Dw|^{2}) {\rm
d}z\nonumber\\[.2pc]
&+ \int_{z_{1}}^{z_{2}} \left( \alpha^{4} - \frac{[
(U'')^{2} ]_{\max} c_{i}}{c_{i}^{3}} - \frac{[
g\beta |U''|]_{\max}}{c_{i}^{3}} \right) |w|^{2} {\rm d}z
\leq 0.
\end{align}
Now, on account of the non-negativity of first integral in (3.20) we get
\begin{equation}
\left( \alpha^{4} - \frac{[ (U'')^{2} ]_{\max} c_{i}}{c_{i}^{3}} -
\frac{[ g\beta|U''|]_{\max}}{c_{i}^{3}} \right) \leq 0
\end{equation}
or
\begin{equation}
\alpha^{3} c_{i}^{3} \leq \frac{[(U'')^{2}]_{\max} c_{i}}{\alpha} +
\frac{[g\beta |U''|]_{\max}}{\alpha}.
\end{equation}
Inequality (3.22) implies that $\alpha c_{i} \rightarrow 0$ as $\alpha
\rightarrow \infty$. This completes the proof of the theorem.
\end{proof}

Here, it is to be noted that for the case $\beta = 0$, from
inequality~(3.22), we get the result of Banerjee {\it et~al} \cite{1}
for the special case of inviscid homogeneous parallel shear flows.


\begin{thebibliography}{99}
\bibitem{1} Banerjee~M~B, Shandil~R~G and Vinay Kanwar, A proof of
Howard's conjecture in homogeneous parallel shear flows, {\it Proc.
Indian Acad. Sci. (Math. Sci)} {\bf 104} (1994)\break 593--595

\bibitem{2} Banerjee~M~B, Shandil~R~G, Prakash~J, Bandral~B~S and Lal
Prem, On Howard's conjecture in heterogeneous shear flows instability of
modified $s$-waves, {\it Indian J. Pure Appl. Math.} {\bf 28(6)} (1997)
825--834

\bibitem{3} Garcia~R~V, Unpublished Lecture Notes (Los Angeles:
University of California) (1961)

\bibitem{4} Howard~L~N, Note on a paper of John W Miles, {\it J. Fluid
Mech.} {\bf 10} (1961) 509--512

\bibitem{5} Miles~J~W, On the stability of heterogeneous shear flows,
{\it J. Fluid Mech.} {\bf 10} (1961)\break 456--508
\end{thebibliography}
\end{document}